# Balmer and Rydberg Equations for Hydrogen Spectra Revisited

**Raji Heyrovska**

Institute of Biophysics, Academy of Sciences of the Czech Republic, 135 Kralovopolska, 612 65 Brno, Czech Republic. Email: rheyrovs@hotmail.com

**Abstract**

Balmer equation for the atomic spectral lines was generalized by Rydberg. Here it is shown that 1) while Bohr's theory explains the Rydberg constant in terms of the ground state energy of the hydrogen atom, quantizing the angular momentum does not explain the Rydberg equation, 2) on reformulating Rydberg's equation, the principal quantum numbers are found to correspond to integral numbers of de Broglie waves and 3) the ground state energy of hydrogen is electromagnetic like that of photons and the frequency of the emitted or absorbed light is the difference in the frequencies of the electromagnetic energy levels.

Subject terms: Chemical physics, Atomic physics, Physical chemistry, Astrophysics

An introduction to the development of the theory of atomic spectra can be found in [1, 2]. This article is divided into several sections and starts with section 1 on Balmer's equation for the observed spectral lines of hydrogen.

**1. Balmer's equation**

Balmer's equation [1, 2] for the observed emitted (or absorbed) wavelengths ($\lambda$) of the spectral lines of hydrogen is given by,

$$\lambda = h_{B,n1}[n_2^2/(n_2^2 - n_1^2)] \tag{1}$$

where $n_1$ and $n_2$ ($> n_1$) are integers with $n_1$ having a constant value, $h_{B,n1} = n_1^2 h_B$ is a constant for each series and $h_B$ is a constant. For the Lyman, Balmer, Paschen, Bracket and Pfund series, $n_1 = 1, 2, 3, 4$ and $5$ respectively. Note: In equation (1), Balmer used different symbols. The symbols used here conform with those below.

For each series, $\lambda$ varies between two limits. When $n_2 \gg n_1$, $\lambda = \lambda_{min,n1} = h_{B,n1} = n_1^2 h_B$ is the minimum wavelength, and it has the values, $h_B$, $4h_B$, $9h_B$, $16h_B$ and $25h_B$ for the Lyman, Balmer, Paschen, Bracket and Pfund series, respectively. Balmer's equation (1) can be written in a new form as,

$$\lambda = \lambda_{min,n1}[n_2^2/(n_2^2 - n_1^2)] = \lambda_{min,n1}/\sin^2\theta \tag{2}$$

where $\sin^2\theta = (n_2^2 - n_1^2)/n_2^2$ in a right angled triangle with $n_2$ as the hypotenuse and $n_1$ as the side which makes an angle $\theta$ with the hypotenuse. $\lambda = \lambda_{min,n1}$ for $\sin^2\theta = 1$.

For the higher wavelength limit, $n_2 = n_1 + 1$ and $\lambda = \lambda_{max,n1}$ is given by

$$\lambda_{max,n1} = \lambda_{min,n1}[(n_1 + 1)^2/(2n_1 +1)] \qquad (3)$$

and is equal to $(4/3)\lambda_{min,n1}$, $(9/5)\lambda_{min,n1}$, $(16/7)\lambda_{min,n1}$, $(25/9)\lambda_{min,n1}$ and $(36/11)\lambda_{min,n1}$, for the Lyman, Balmer, Paschen, Bracket and Pfund series, respectively.

The wavelength data from [3] for the various spectral lines of hydrogen are given in Table 1. Fig. 1 shows the linear graphs of $\lambda$ versus $\sin^{-2}\theta$. The slopes give the limiting values, $\lambda_{min,n1} = h_{B,n1} = n_1^2 h_B$. These values are also included in Table 1 and shown as the intercepts in Fig. 1. For the Pfund series ($n_1 = 5$), $\lambda_{min,n1} = n_1^2 h_B$, where $h_B = \lambda_{min,n1}$ is the value for the Lyman series ($n_1 = 1$).

**2. Rydberg equation**

Balmer's equation (1) was generalized by Rydberg as shown [1, 2],

$$1/\lambda = R_H[(1/n_1^2) - (1/n_2^2)] \qquad (4)$$

where $1/\lambda$ is the wave number of the emitted (or absorbed) light and $R_H = 1/h_B = n_1^2/h_{B,n1} = n_1^2/\lambda_{min,n1}$ is the Rydberg's constant. For each series, $1/\lambda$ varies linearly with $(1/n_2)^2$, with $R_H$ (= $1/h_B$) as the slope and $R_H/n_1^2$ (= $1/h_{B,n1} = 1/\lambda_{min,n1}$) as the intercept. The interpretation of the Rydberg equation (4) amounts to that of the Rydberg constant, $R_H$ and of the term in the square brackets, $[(1/n_1^2) - (1/n_2^2)]$.

**3. Rydberg constant ($R_H$) and the energy ($E_H$) of hydrogen**

Bohr [4] made a major advance in the interpretation of the Rydberg constant [1, 2]. On multiplying both sides of equation (4) by hc, where c is the speed of light in vacuum and h is the Planck's constant, the energy, hν, of emitted (or absorbed) light is given by,

$$hc/\lambda = h\nu = E_H[(1/n_1)^2 - (1/n_2)^2] = E_{H,n1} - E_{H,n2} \quad (5)$$

where $\nu = c/\lambda$ is the frequency of the emitted (or absorbed) light, $E_H = hcR_H$ is the ground state energy of the hydrogen atom, $E_{H,n1} = E_H/n_1^2$ and $E_{H,n2} = E_H/n_1^2$.

$E_H = eI_H$, where e is the unit of charge and $I_H$ is the measured ionization potential and it is given by the equivalent terms, [1, 2, 4],

$$E_H = (1/2)\mu_H v_H^2 = (1/2)p_\omega \omega_H \quad (6)$$

where $\mu_H$ is the reduced mass of the hydrogen atom, $v_H = \omega_H a_H$ is the linear velocity, $\omega_H$ is the angular velocity, $a_H$ is the Bohr radius (distance between the electron and proton), $p_\omega$ is angular momentum. These constants are given by the following equations,

$$p_\omega = \mu_H v_H a_H = h/2\pi \quad (7a)$$
$$\mu_H v_H \lambda_{dB,H} = h \quad (7b)$$
$$I_H = (e/8\pi\varepsilon_o a_H) \quad (7c)$$

where $4\pi\varepsilon_o$ is the electrical permittivity, $\varepsilon_o$ is the electric constant and $\lambda_{dB,H} = 2\pi a_H$ is the de Broglie wavelength [1, 2]. The values of the above various constants can be found in [5, 6, 7]. Through equation (6), Bohr's theory accounts well for the Rydberg constant [6], $R_H = E_H/hc$.

**4. Why quantization of angular momentum is unsatisfactory**

The significance of the term with the reciprocals of squares of integers in the square brackets in the equations (4) and (5) puzzled many scientists. In order to explain this, Bohr empirically assumed quantization of the angular momentum, $p_{\omega,n}$

$$p_{\omega,n} = nh/2\pi = n\mu_H v_H a_H \qquad (8)$$

where n is called the principle quantum number. Equation (8) implies that $v_{H,n} = nv_H$ and therefore, equation (6) becomes for any n,

$$E_{H,n} = (1/2)\mu_H v_{H,n}^2 = n^2 E_H = n^2 eI_H \qquad (9)$$

where $E_H$ is the value of $E_{H,n}$ for n = 1 (ground state). Thus, $E_{H,n}$ increases with $n^2$ and the energy difference,

$$E_{H,n1} - E_{H,n2} = E_H(n_1^2 - n_2^2) \qquad (10)$$

does not agree with equation (5). Hence, the quantization of the angular momentum as in equation (8) does not explain the Rydberg equation (4). The author came to a similar conclusion in an earlier paper [7]. A recent review [8] of the literature around the period shows that many had objections to Bohr's theory.

**5. An alternate more meaningful form of the Rydberg equation**

Here, the difference in the squares of numbers in equations (1) and (4) is interpreted in a different way. As in equation (2), the Rydberg equation (4) can be written as,

$$1/\lambda = (1/\lambda_{min,n1})[n_2^2/(n_2^2 - n_1^2)] = (1/\lambda_{min,n1})\sin^2\theta \qquad (11)$$

where $(n_2^2 - n_1^2)/n_2^2 = \sin^2\theta$. The wave numbers $1/\lambda$ for any given series increase directly with $\sin^2\theta$ from a minimum value for $n_2 = n_1 + 1$, to a maximum corresponding to $\sin^2\theta = 1$ for $n_2 \gg n_1$. Figure 2 shows the linear graphs of $1/\lambda$ versus $\sin^2\theta$ for the wavelength data in [3]. The slopes of these lines give the values of $R_H/n_1^2 = 1/\hbar_{B,n1} = 1/\lambda_{min,n1}$.

On multiplying both sides of equation (11) by hc, and using Bohr's equality, $R_H = E_H/hc$, the following relations are obtained,

$$h\nu = E_{H,n1} - E_{H,n2} = (hcR_H/n_1^2)\sin^2\theta = E_{H,n1}\sin^2\theta \qquad (12)$$

$$E_{H,n1} = E_H/n_1^2 = (1/2)\mu_H v_1^2 \qquad (13)$$

$$h\nu = (1/2)\mu_H(v_1\sin\theta)^2 = (1/2)\mu_H(v_1^2 - v_2^2) \qquad (14)$$

where $v_1 = v_H/n_1$ and $v_2 = v_H/n_2$. Equation (14) shows that emission (or absorption) of light changes the velocity $v_1$ by $v_1\sin\theta$ to $v_2$.

## 6. Principal quantum numbers as the number of de Broglie waves

For $v_1 = v_H/n_1$, the angular momentum and de Broglie wavelengths are given by,

$$p_{\omega,1} = p_\omega/n_1 = \mu_H v_1 a_H = h/2\pi n_1 \qquad (15)$$

$$h/\mu_H = v_H \lambda_{dB,H} = v_1 \lambda_{dB,n1} = \text{constant} \qquad (16)$$

From equation (15), $p_{\omega,1} = p_\omega/n_1 = h/2\pi n_1$ unlike in Bohr's equation (8) where $p_{\omega,n} = nh/2\pi$. Equation (16) shows that since $h/\mu_H$ is a constant, the product $v_H\lambda_{dB,H}$ is a constant. The de Broglie wavelength, $\lambda_{dB,H} = 2\pi a_H$ for $n_1 = 1$ is the circumference of a circle with radius $a_H$ and $\lambda_{dB,n1} = n_1\lambda_{dB,H}$ is the de Broglie wavelength of $n_1$ integral number of de Broglie waves and is the circumference of a circle with radius $n_1 a_H$. Thus, the following relations are obtained,

$$h\nu = E_{H,n1}\sin^2\theta = (1/2)\mu_H(v_1\sin\theta)^2 = (1/2)(h/\mu_H\lambda_{dB,H})^2[(1/n_1)^2 - (1/n_2)^2] \quad (17a)$$

$$E_{H,n1} = E_H/n_1^2 = (1/2)\mu_H v_1^2 = (1/2)(h/\mu_H\lambda_{dB,H,n1})^2 \quad (17b)$$

$E_{H,n1}$ decreases linearly with increase in $(n_1\lambda_{dB,H})^2$.

Thus, the principal quantum numbers $n_1$, $n_2$ etc represent the number of standing de Broglie waves.

### 7. Energy of hydrogen $E_H$ interpreted as $h\nu_{max}$ as for photons

It is shown here that on combining the above results with equation (2),

$$E_{H,n1} = E_H/n_1^2 = hcR_H/n_1^2 = hc/\lambda_{min,n1} = h\nu_{max,n1} \quad (18a)$$

$$E_{H,n2} = E_H/n_2^2 = hcR_H/n_2^2 = hc/h_{B,n2} = hc/\lambda_{min,n2} = h\nu_{max,n2} \quad (18b)$$

For the Lyman series, $n_1 = 1$ and $cR_H = c/\lambda_{min,H} = \nu_{max,H}$ the frequency of light. It therefore follows that the energy of hydrogen $E_H$ given by equation (6) is equal to,

$$E_H = h\nu_{max,H} = (1/2)\mu_H v_H^2 = eI_H \quad (19)$$

as for photons. On combining with the earlier result in [5],

$$E_H = eI_H = (1/2)e^2/C_B = h\nu_{max,H} \qquad (20)$$

is the electromagnetic energy of a condenser with unit opposite charges separated by a distance of $a_H$, where $C_B = 4\pi\varepsilon_o a_H$ is the Golden mean capacity [5].

The frequency, $\nu = c/\lambda$ of the emitted (or absorbed) light for any series corresponding to the equation (5) is therefore the difference in the frequencies,

$$\nu = (E_{H,n1} - E_{H,n2})/h = \nu_{max,H}[(1/n_1)^2 - (1/n_2)^2] = (\nu_{max,n1} - \nu_{max,n2}) \qquad (21)$$

where, the wave number, $\nu_{max,H}/c = 1/\lambda_{min,H} = R_H$, the Rydberg constant. The values of $h\nu$ are given in the last column in Table 1.

**Acknowledgement:** The author thanks the Institute of Biophysics of the Academy of Sciences of the Czech Republic for the financial support.

**References**

1. White, H.E.: Introduction to Atomic Spectra. McGraw-Hill, New York (1934)

2. a) http://www.wadsworthmedia.com/marketing/sample_chapters/0534266584_ch09.pdf

c) http://hyperphysics.phy-astr.gsu.edu/hbase/hyde.html

b) http://en.wikipedia.org/wiki/Book:Hydrogen

3. http://hyperphysics.phy-astr.gsu.edu/Hbase/tables/hydspec.html#c1; data from: Blatt, F.J.: Modern Physics. McGraw-Hill, New York (1992)

4. Bohr, N.: Philos. Mag. **26**, 1 (1913);

http://web.ihep.su/dbserv/compas/src/bohr13/eng.pdf

5. Heyrovska, R.: Mol. Phys. **103**, 877 (2005).

6. http://physics.nist.gov/cuu/Constants/

7. Heyrovska, R.: in Frontiers of Fundamental Physics, Vol. 3, ed. B. G. Sidharth; Proceedings of the Fifth International Conference on Frontiers in Fundamental Physics, Hyderabad, January 2003. Universities Press (Division of Orient Longman), 2007, chapter 8, pp. 203 - 215

8. H. Kragh, http://ivs.au.dk/fileadmin/www.ivs.au.dk/reposs/reposs-009.pdf

RePoSS: Research Publications on Science Studies, RePoSS #9: July 2010

**Table 1. Wavelength data [3], etc for the spectral lines of atomic hydrogen**

| Wavelength λ (nm) | Relative Intensity | Transition | Color or region of EM spectrum | $n_2$ | $\sin^{-2}\theta$ | $1/\lambda$ (nm$^{-1}$) | $\sin^2\theta$ | $\nu$ ($10^{15}$ s$^{-1}$) | $h\nu$ ($10^{-18}$ J) |
|---|---|---|---|---|---|---|---|---|---|
| **91.170** | \multicolumn{3}{Lymann Series ($n_1 = 1$)} | | **1.000** | **0.01097** | **1.000** | **3.2883** | **2.1788** |
| 93.782 | ... | 6 -> 1 | UV | 6 | 1.029 | 0.01066 | 0.972 | 3.1967 | 2.1182 |
| 94.976 | ... | 5 -> 1 | UV | 5 | 1.042 | 0.01053 | 0.960 | 3.1565 | 2.0915 |
| 97.254 | ... | 4 -> 1 | UV | 4 | 1.067 | 0.01028 | 0.938 | 3.0826 | 2.0425 |
| 102.583 | ... | 3 -> 1 | UV | 3 | 1.125 | 0.00975 | 0.889 | 2.9224 | 1.9364 |
| 121.566 | ... | 2 -> 1 | UV | 2 | 1.333 | 0.00823 | 0.750 | 2.4661 | 1.6340 |
| **364.600** | | Balmer Series ($n_1 = 2$) | | | **1.000** | **0.00274** | **1.000** | **0.8223** | **0.5448** |
| 383.538 | 5 | 9 -> 2 | Violet | 9 | 1.052 | 0.00261 | 0.951 | 0.7816 | 0.5179 |
| 388.905 | 6 | 8 -> 2 | Violet | 8 | 1.067 | 0.00257 | 0.938 | 0.7709 | 0.5108 |
| 397.007 | 8 | 7 -> 2 | Violet | 7 | 1.089 | 0.00252 | 0.918 | 0.7551 | 0.5004 |
| 410.174 | 15 | 6 -> 2 | Violet | 6 | 1.125 | 0.00244 | 0.889 | 0.7309 | 0.4843 |
| 434.047 | 30 | 5 -> 2 | Violet | 5 | 1.190 | 0.00230 | 0.840 | 0.6907 | 0.4577 |
| 486.133 | 80 | 4 -> 2 | Bluegreen (cyan) | 4 | 1.333 | 0.00206 | 0.750 | 0.6167 | 0.4086 |
| 656.272 | 120 | 3 -> 2 | Red | 3 | 1.800 | 0.00152 | 0.556 | 0.4568 | 0.3027 |
| 656.285 | 180 | 3 -> 2 | Red | 3 | 1.800 | 0.00152 | 0.556 | 0.4568 | 0.3027 |
| **820.250** | | Paschen Series ($n_1 = 3$) | | | **1.000** | **0.00122** | **1.000** | **0.3655** | **0.2422** |
| 954.620 | ... | 8 -> 3 | IR | 8 | 1.164 | 0.00105 | 0.859 | 0.3140 | 0.2081 |
| 1004.980 | ... | 7 -> 3 | IR | 7 | 1.225 | 0.00100 | 0.816 | 0.2983 | 0.1977 |
| 1093.800 | ... | 6 -> 3 | IR | 6 | 1.333 | 0.00091 | 0.750 | 0.2741 | 0.1816 |
| 1281.810 | ... | 5 -> 3 | IR | 5 | 1.563 | 0.00078 | 0.640 | 0.2339 | 0.1550 |
| 1875.010 | ... | 4 -> 3 | IR | 4 | 2.286 | 0.00053 | 0.438 | 0.1599 | 0.1059 |
| **1452.270** | | Brackett Series ($n_1 = 4$) | | | **1.000** | **0.00069** | **1.000** | **0.2064** | **0.1368** |
| 2630.000 | ... | 6 -> 4 | IR | 6 | 1.800 | 0.00038 | 0.556 | 0.1140 | 0.0755 |
| 4050.000 | ... | 5 -> 4 | IR | 5 | 2.778 | 0.00025 | 0.360 | 0.0740 | 0.0490 |
| **2279.250** | | Pfund Series ($n_1 = 5$) | | | **1.000** | **0.00044** | **1.000** | **0.1315** | **0.0872** |
| 7400.000 | ... | 6 -> 5 | IR | 6 | 3.273 | 0.00014 | 0.306 | 0.0405 | 0.0268 |

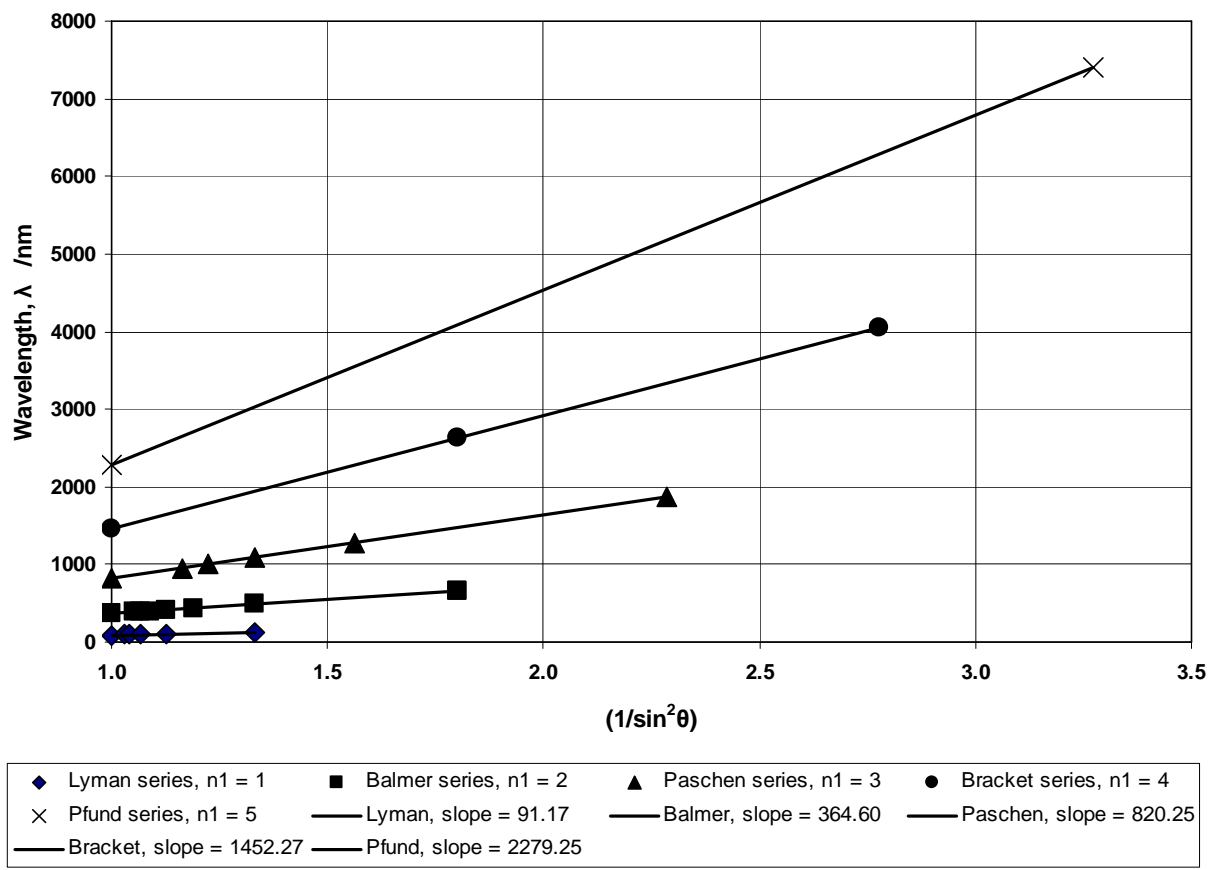

**Figure 1.**

Linear graphs of wavelength λ versus $(1/\sin^2\theta)$ for the spectral lines of hydrogen according to the new form of Balmer equation (2). The slopes of the lines are equal to the limiting values, $\lambda_{min,n1} = h_{B,n1} = n_1^2 h_B$. Data for λ are from [3].

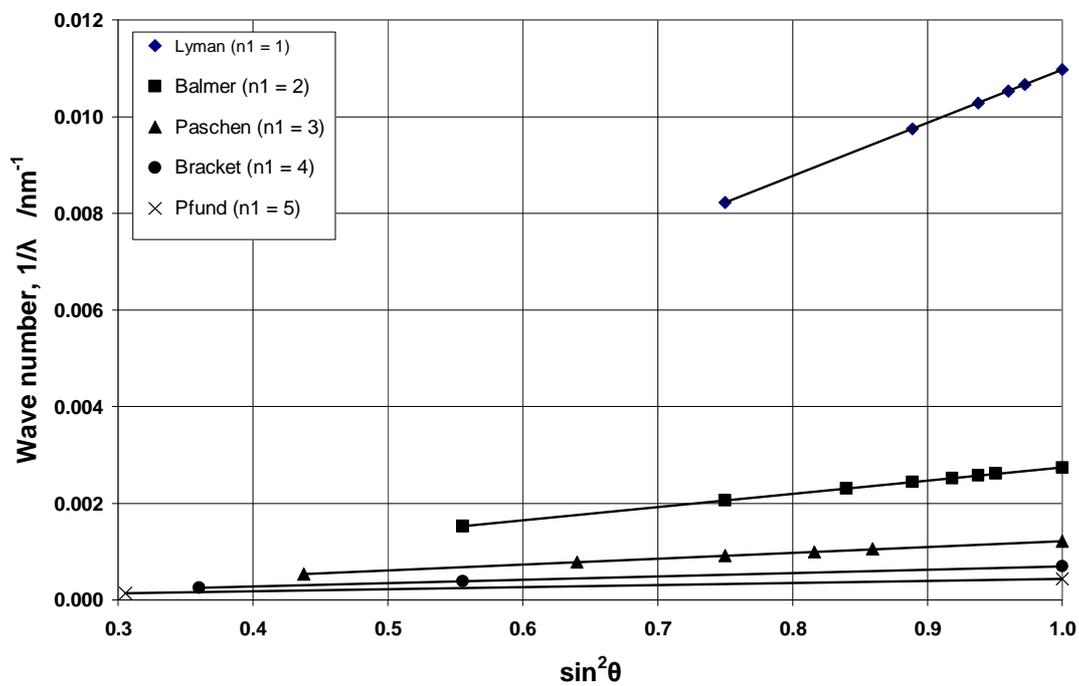

**Figure 2.**

Linear graphs of wave numbers (1/λ) versus $\sin^2\theta$ for the spectral lines of hydrogen according to the new concise alternate form of the Rydberg equation (11). The slopes of the lines are equal to $R_H/n_1^2$, where $R_H = E_H/hc$. Data for λ are from [3].